\documentclass[alpha-refs]{wiley-article}

\usepackage{siunitx}
\usepackage{booktabs}
\usepackage{multirow}

\usepackage[most]{tcolorbox}
\newtcolorbox[]{Practitionernotes}[1][]{%
    enhanced,
    breakable,
    colback=white,
    colbacktitle=white,
    coltitle=black,
    fonttitle=\bfseries,
    boxrule=.6pt,
    titlerule=.2pt,
    toptitle=3pt,
    bottomtitle=3pt,
    title=Practitioner notes,
    #1}

\papertype{Original Article}

\title{Beyond Self-Regulated Learning Processes: Unveiling Hidden Tactics in Generative AI-Assisted Writing}

\author[1\authfn{1}]{Kaixun Yang}
\author[1,2\authfn{1}]{Yizhou Fan}
\author[2]{Luzhen Tang}
\author[1]{Mladen Rakovi\'{c}}
\author[1]{Xinyu Li}
\author[1]{Dragan Ga\v{s}evi\'{c}}
\author[1]{Guanliang Chen}

\contrib[\authfn{1}]{Equally contributing authors.}

\affil[1]{Centre for Learning Analytics, Faculty of Information Technology, Monash University, Clayton, Victoria 3800, Australia}
\affil[2]{Graduate School of Education, Peking University, Beijing, 100871, China}

\corraddress{Guanliang Chen, Centre for Learning Analytics, Faculty of Information Technology, Monash University, Clayton, Victoria 3800, Australia}
\corremail{Guanliang.Chen@monash.edu}

\fundinginfo{National Natural Science Foundation of China, Grant/Award Number: 62407001.}

\runningauthor{YANG ET AL.}

\begin{document}

\begin{frontmatter}
\maketitle

\begin{abstract}
The integration of Generative AI (GenAI) into education is reshaping how students learn, making self-regulated learning (SRL) — the ability to plan, monitor, and adapt one’s learning — more important than ever. To support learners in these new contexts, it is essential to understand how SRL unfolds during interaction with GenAI tools. Learning analytics offers powerful techniques for analyzing digital trace data to infer SRL behaviors. However, existing approaches often assume SRL processes are linear, segmented, and non-overlapping—assumptions that overlook the dynamic, recursive, and non-linear nature of real-world learning. We address this by conceptualizing SRL as a layered system: observable learning patterns reflect hidden tactics (short, purposeful action states), which combine into broader SRL strategies. Using Hidden Markov Models (HMMs), we analyzed trace data from higher education students engaged in GenAI-assisted academic writing. We identified three distinct groups of learners, each characterized by different SRL strategies. These groups showed significant differences in performance, indicating that students' use of different SRL strategies in GenAI-assisted writing led to varying task outcomes. Our findings advance the methodological toolkit for modeling SRL and inform the design of adaptive learning technologies that more effectively support learners in GenAI-enhanced educational environments.

\end{abstract}
\end{frontmatter}

\begin{Practitionernotes}
What is already known about this topic
\begin{itemize}
    \item Generative artificial intelligence (GenAI) is being increasingly integrated into education to enhance human learning.
    \item Self-regulated learning (SRL) is recognized as a crucial set of competencies to ensure academic success and productive lifelong learning.
    \item Understanding how SRL enact their learning strategies during a task has been on the researchers’ agenda for many years.
\end{itemize}

What this paper adds
\begin{itemize}
    \item We found that hidden learning tactics can be effectively captured as the latent states in Hidden Markov Models.
    \item We found that identifying learning strategies from hidden learning tactics can provide a more accurate depiction of the discontinuous, non-linear, and interconnected nature of SRL.
    \item Based on the identified learning strategies, we found that students who performed significantly better on tasks tended to engage more with GenAI and less in cognitive and metacognitive activities.
\end{itemize}

Implications for practice and/or policy
\begin{itemize}
    \item Practitioners should be aware of students' potential over-reliance on GenAI techniques, which may lead them to miss valuable opportunities for deep, meaningful learning.
    \item Researchers should recognize that modeling SRL with careful attention to its complex nature can yield more nuanced insights.
\end{itemize}
\end{Practitionernotes}

\section{INTRODUCTION}
Generative artificial intelligence (GenAI) is being increasingly integrated into education to enhance human learning \citep{yan2024promises}. Although GenAI can automate many aspects of the learning process \citep{lee2022coauthor,naseer2024automated,tate2024can}, it is important to emphasize that AI should augment human intellect and capabilities rather than replace them \citep{akata2020research}. As a result, learners’ regulation becomes crucial, serving as a fundamental mechanism in an individual’s ability to engage effectively in learning \citep{taranto2020sustaining} — especially in the age of GenAI. Self-regulated learning (SRL) is a cyclical, constructive, and active process through which learners engage with learning tasks by enacting cognitive, metacognitive, motivational, and emotional  processes \citep{schunk2011handbook,azevedo2015defining}. SRL is recognized as a crucial set of competencies to ensure academic success \citep{dent2016relation} and productive lifelong learning \citep{nguyen2024lifelong}. Understanding how SRL unfolds during interactions with GenAI tools is crucial for better supporting learners in these new contexts.

An essential aspect of productive SRL is developing and enacting a repertoire of SRL strategies, which are sequences of learning processes that learners purposefully engage to achieve their learning goals \citep{hacker1998metacognition,srivastava2022effects}. 
Understanding how self-regulated learners enact their learning strategies during a task has been on the researchers' agenda for many years \citep{villalobos2024exploring, jovanovic2017learning,saqr2023intense,gasevic2017detecting,rakovic2024conversation}, as such an understanding enables researchers to uncover the cognitive and metacognitive mechanisms that drive SRL, offering insight into the internal processes learners use to regulate their learning \citep{zimmerman2000attaining,winne2017learning,winne2022modeling}. Furthermore, identifying learning strategies allows learning technologies to adapt content and scaffolding in real time, providing personalized support to meet individual learner needs \citep{roll2011improving,azevedo2013international,lim2024students,molenaar2022concept}.

\begin{figure*}[hbt!]
\centering
\includegraphics[width=1\textwidth]{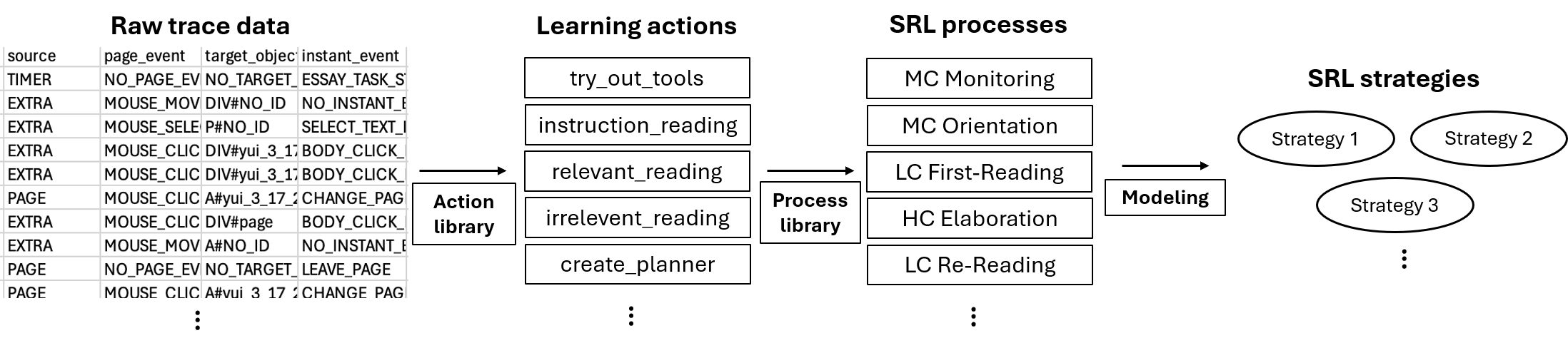}
\caption{SRL strategies modeling procedure proposed by \cite{fan2022towards}. SRL theoretical model adapted from \cite{bannert2007metakognition}. MC, HC, and LC represent metacognition, high cognition, and low cognition.} 
\label{fig1}
\end{figure*}

Recognizing the significance of SRL strategies, the learning analytics community has shown growing interest in developing innovative analytical methods to detect them. In particular, researchers are leveraging fine-grained trace-data from digital learning platforms to identify the SRL strategies learners enact, as trace data can capture how SRL processes dynamically unfold during learning \citep{matcha2019analytics,saint2020combining,fan2021learning,srivastava2022effects,li2024analytics,cheng2025self}. In a widely used framework proposed by \cite{fan2022towards} (see Figure \ref{fig1}), trace data (e.g., mouse-click events) generated from students’ interactions with digital tools during learning tasks are first mapped to learning actions (e.g., instruction reading) using a predefined action library. These learning actions are then mapped to SRL processes (e.g., metacognitive-planning) based on theoretical SRL frameworks via a predefined process library. Finally, these SRL processes are modeled to infer the underlying SRL strategies. Based on this framework, multiple modeling methods have been proposed to identify SRL strategies \citep{srivastava2022effects,li2024analytics,alghamdi2025analytics,cheng2025self,fan2023dissecting}. For instance, \cite{srivastava2022effects} proposed using a First-Order Markov Model (FOMM) and the Expectation-Maximization clustering algorithm to identify SRL strategies in multi-source writing tasks from SRL processes, successfully uncovering three theoretically meaningful SRL strategies. Similarly, \cite{cheng2025self} employed Epistemic Network Analysis (ENA) to identify SRL strategies used by high-performing and low-performing student groups based on their SRL processes in multi-source writing.

While trace-based analytical methods have shown considerable promise in identifying and interpreting SRL strategies, a recurring limitation persists across much of this research. These modeling approaches often rely on rigid and sometimes unrealistic assumptions about the nature of SRL processes. Specifically, they tend to assume that SRL processes are continuous, non-overlapping, and absolutely clear. For instance, the FOMM presupposes that each SRL process depends solely on the immediately preceding process, ignoring the broader historical context. This simplification makes FOMM ill-suited for handling noisy or incomplete data, where historical context may be critical for accurate interpretation. Similarly, ENA is constrained by its reliance on continuous co-occurrence patterns. As a result, it struggles to detect or represent discontinuous or temporally dispersed patterns of SRL process, limiting its ability to capture the complexity and non-linearity of SRL strategies. As highlighted by \cite{dever2023complex}, who explored SRL from the perspective of complexity science, SRL processes are dynamic and non-linear. These processes involve ongoing interactions among cognitive, metacognitive, motivational, and emotional elements that evolve over time. This view is also supported by \cite{li2020examining}, who observed that learners' SRL patterns exhibit non-linearity and are shaped by multiple factors, such as task complexity and individual differences. Additionally, the involvement of GenAI further complicates the SRL processes, as its use may affect learners' SRL behaviors \citep{fan2025beware}. As a result, directly modeling SRL strategies from SRL processes may not adequately capture their discontinuous, non-linear, and intertwined nature, especially in the context of GenAI-assisted learning.

To address the above gap, we argued that latent states should exist between SRL processes and SRL strategies to better reflect the discontinuous, non-linear, and intertwined nature of SRL. These latent states, which we refer to as \texttt{hidden tactics}, represent the unobservable tactics that describe the relationships between students' SRL processes. Incorporating \texttt{hidden tactics} can help identify SRL strategies that more accurately align with real-world learning scenarios. We proposed using the Hidden Markov Model (HMM) to uncover the \texttt{hidden tactics} underlying SRL processes. HMMs are statistical models that represent systems governed by a Markov process with hidden (unobservable) states. These hidden states introduce an additional layer of abstraction, enabling the model to capture more temporally extended dependencies \citep{thyer2003hidden}. Moreover, HMMs include probabilistic emission distributions, allowing each hidden state to produce a variety of possible observations with associated probabilities \citep{ephraim2002hidden}. This feature makes HMMs well-suited for handling uncertainty and noise in observed SRL processes. HMMs have been used in previous SRL studies \citep{fincham2018study,villalobos2024exploring}, where they were applied to raw trace data. However, using HMMs on raw trace data makes it difficult to infer the self-regulatory processes that underpin tactics and strategies \citep{fan2021learning}. In contrast, we applied HMMs to processed trace data, which first identified SRL processes, enabling a clearer extraction of tactics and strategies.

To investigate the SRL strategies learners employ during writing tasks involving GenAI, we applied the proposed method to a dataset comprising students in higher education who used GenAI assistance for writing collected via the FLoRA system \citep{li2024flora}. We also compared our findings with those derived from a traditional benchmark method \citep{srivastava2022effects} that does not account for \texttt{hidden tactics}. We identified nine hidden tactics. Based on these, we uncovered three distinct clusters of students employing different SRL strategies. Furthermore, we found statistically significant differences in task performance across these groups, which were not detected by the benchmark method, demonstrating the discriminatory power of our approach. Formally, our research addresses the following \textbf{R}esearch \textbf{Q}uestions:

\begin{enumerate}[label=\bfseries RQ\arabic*,leftmargin = 30pt]
    \item What are the common hidden tactics behind SRL processes in GenAI-assisted writing tasks?
    \item What are the common SRL strategies that can be detected by taking students’ SRL hidden tactics into consideration?
    \item To what extent do students' SRL strategies based on SRL hidden tactics relate to task performance?
\end{enumerate}

\section{BACKGROUND}
\subsection{Concepts in SRL modeling}
In the SRL modeling procedure illustrated in Figure \ref{fig1}, it is important to distinguish between the following concepts: trace data, learning actions, SRL processes, and SRL strategies.

Trace data are the digital footprints learners leave behind as they interact with online platforms. These typically consist of time-stamped, system-generated logs that capture behaviors such as clickstreams, time-on-task, page views, and multimodal data like eye-tracking \citep{winne2020construct}. While trace data are inherently low-level and granular, and thus not directly explanatory, they provide a valuable proxy and foundational evidence for inferring higher-order cognitive and metacognitive processes \citep{biswas2017data,jarvela2019capturing,molenaar2023measuring}.

To make trace data interpretable, researchers often translate raw logs into learning actions, which are discrete and contextually grounded units of behavior that reflect the pedagogical structure of a learning task \citep{bannert2014process,molenaar2023measuring}. For instance, during a writing task, continuous keystroke entries may be characterized as essay writing, whereas accessing a page containing task requirements may be interpreted as reading task instructions. These learning actions provide a meaningful middle layer between raw data and the theoretical constructs of SRL \citep{fan2022towards}.

Beyond isolated individual learning actions, SRL processes involve broader patterns of regulation that unfold over time. Several theoretical models and frameworks have been proposed \citep{inbook,bannert2007metakognition,zimmerman2013theories,winne2018theorizing} to explain SRL. One such model is COPES introduced by \cite{winne1998studying} and further elaborated by \cite{winne2017learning,winne2018theorizing}, which outlines four key phases in the learning process: task definition, goal setting and planning, enacting study tactics and strategies, and adapting studying. Each phase is described in terms of five learning facets—conditions, operations, products, evaluations, and standards (COPES) — and incorporates various cognitive and metacognitive processes such as monitoring, searching, and evaluation. Similarly, \cite{bannert2007metakognition} offers a model that categorizes SRL into cognitive, metacognitive, and motivational processes. Expanding on this, \cite{fan2022towards} further operationalized Bannert’s model by defining cognitive and metacognitive processes as sequences of learning actions. These comprise four metacognitive processes (i.e., orientation, planning, monitoring, and evaluation) and three cognitive processes (i.e., first reading, re-reading, and elaboration/organization). For example, when the learner clicked on the timer (trace), this action was labeled as TIMER (learning action) and interpreted as MetaCoginition.Monitoring (SRL process), as it reflects the learner’s monitoring of the remaining time to complete the task. Such frameworks help researchers align observations based on trace-data with deeper psychological mechanisms of regulation.

Over the past decade, SRL strategies have primarily been identified through the analysis of sequences of learning actions captured in trace data (i.e., sequences of learners’ interactions with digital learning resources) \citep{maldonado2018mining}. However, some critics argue that interpreting strategies based solely on learning actions is highly task- and context-dependent \citep{fan2021learning}, making it challenging to compare strategies across different settings \citep{srivastava2022effects}. In response, recent studies have focused on interpreting these sequences of learning actions as theoretically grounded SRL processes \citep{fan2021learning}. These SRL processes exist regardless of the type of task (e.g., monitoring can occur in both writing and reading tasks), whereas some learning actions may not be present in certain tasks (e.g., writing actions are unlikely in reading tasks). This shift allows for the detection of SRL strategies from SRL processes that can be applied across various contexts \citep{srivastava2022effects,li2024analytics,cheng2025self}.

However, previous studies have shown that SRL processes are often discontinuous, non-linear, and intertwined, rather than following a fixed, step-by-step sequence \citep{li2020examining,dever2023complex}. For example, during a writing task, a student may realize their arguments are weak while writing (i.e., Elaboration/Organization), so they might return to the reading materials (i.e., Re-reading) and simultaneously revise their goals (i.e., Planning). Such overlapping and dynamic behaviors illustrate the complexity of SRL and pose challenges for approaches that attempt to map strategies directly from observed SRL processes. To address this, we propose introducing an intermediate layer of analysis: \texttt{hidden tactics}, which helps preserve the nature of SRL processes by modeling tactics that reflects the probability distribution of SRL processes when converting them into SRL strategies. To the best of our knowledge, this is the first study to operationalize \texttt{hidden tactics} as an analytical construct for bridging SRL processes and strategies. We also validated our approach by comparing it against previous benchmark methods. 

\subsection{Studies on GenAI-assisted writing}
In the early stages, research primarily focused on developing systems to support GenAI-assisted writing \citep{lee2022coauthor, jakesch2023co}. For example, \cite{lee2022coauthor} developed a GPT-3-based system that provides sentence-level suggestions to writers during the writing process. Similarly, \cite{jakesch2023co} introduced a tool that generates text continuations based on user input. Building on these systems, several studies began exploring patterns that emerge during GenAI-assisted writing \citep{shibani2023visual,nguyen2024human}. These patterns have been linked to various aspects such as writing performance \citep{yang2025modifying} and cognitive processes \citep{cheng2024evidencecentered, liu2024investigating, yang2024ink}. For instance, \cite{yang2025modifying} used causal modeling on a GenAI-assisted writing dataset and found that active engagement in higher-order cognitive processes was associated with higher essay quality.

Learners’ regulation is a fundamental mechanism for effective engagement in learning, particularly in the age of GenAI \citep{taranto2020sustaining}. As a result, several studies have begun to explore SRL processes in GenAI-assisted writing tasks \citep{fan2025beware,chen2025unpacking}. For example, \cite{chen2025unpacking} investigated how learners’ SRL help-seeking process differs when interacting with GenAI versus human experts during writing tasks. They found that the GenAI group exhibited a nonlinear help-seeking pattern—such as skipping the evaluation phase—which contrasted with the more linear process observed in the human expert group. Similarly, \cite{fan2025beware} conducted a randomized experimental study comparing learners’ motivation, SRL processes, and learning outcomes across groups supported by different agents (i.e., GenAI, human experts, and writing analytics tools). The results showed that while the GenAI group achieved greater improvements in essay scores, their knowledge gains and ability to transfer knowledge did not differ significantly from the other groups.

While significant work has been done, it often falls short in two key areas: (i) many studies overlook the SRL processes involved in GenAI-assisted writing tasks \citep{shibani2023visual,nguyen2024human,yang2025modifying,liu2024investigating}, limiting insight into the deeper psychological mechanisms that drive learning regulation; and (ii) they fail to capture the complex nature of SRL processes (i.e., their intertwined, discontinuous, and non-linear characteristics) \citep{fan2025beware,chen2025unpacking}, thereby restricting their applicability to real-world learning contexts. Our approach addresses these gaps by introducing a novel method that models hidden tactics using HMMs. This allows us to interpret the hidden tactic as a probability distribution over different SRL processes, enabling the identification of SRL strategies and capturing the complexity of SRL.

\section{METHODS}
\subsection{Data}
The data for this study were collected from an English academic writing course at a Chinese university. We designed a reading-writing task based on an online learning platform (i.e., FLoRA system \citep{li2024flora}). The platform provided reading materials related to the task, covering topics such as artificial intelligence, differentiated instruction, and scaffolding teaching. Students were required to write an essay on the future of education based on these materials. Additionally, the platform offered detailed grading criteria for students and integrated a ChatGPT 4.0 interface to facilitate reading and writing during the task. This integration also allowed us to collect the data about the students's use of ChatGPT during the task. A total of 241 students from various disciplines, including chemistry, biology, physics, geography, and computer science, participated in this study (55\% male, mean age = 22.40, SD = 1.01). These participants were non-native English speakers. Notably, over 86\% of the participants had prior experience using ChatGPT before the study began. We recorded the students’ complete learning trace data on the learning platform, including page visit logs, clickstream data, mouse movement, and keyboard input. Additionally, we saved all the essays submitted by the students. To ensure the reliability of the essay grading, we randomly selected 20 essays and had them independently graded by two researchers based on the provided grading criteria. Inter-rater reliability was tested using the intraclass correlation coefficient (ICC) \citep{bartko1966intraclass} and the absolute consistency index, which showed high consistency (all ICC values > 0.85). Given the excellent inter-rater reliability, the remaining essays were graded by a single researcher.

\subsection{Data analysis}
\begin{figure*}[hbt!]
\centering
\includegraphics[width=1\textwidth]{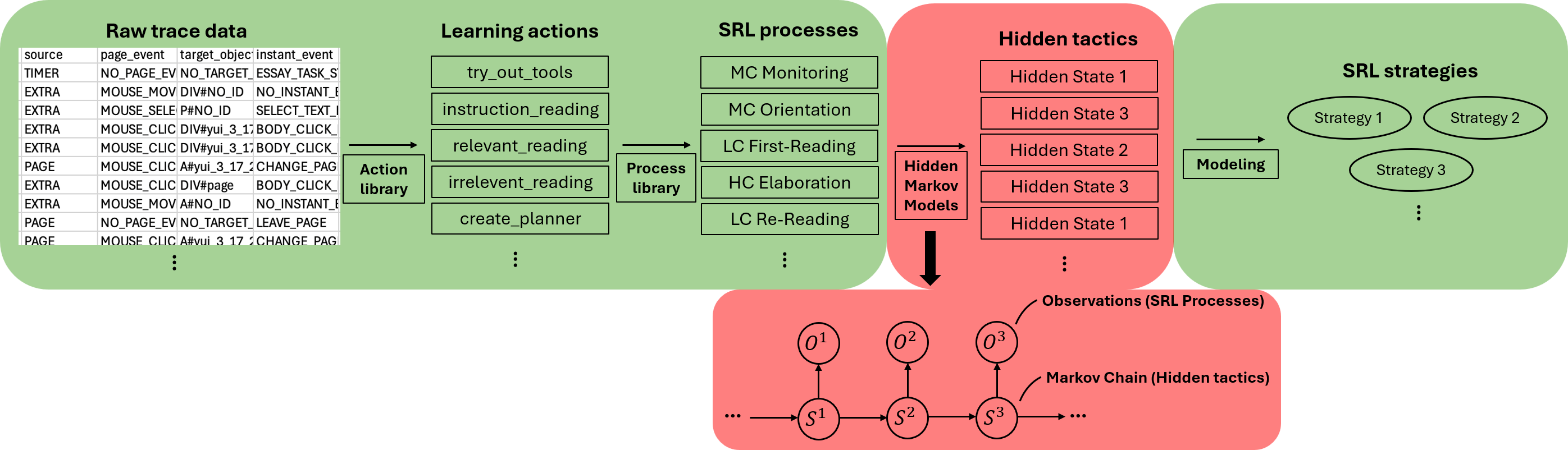}
\caption{Methodology Overview. Elements highlighted with a green background are based on the approach proposed by \cite{fan2022towards}, while those with a red background represent our proposed additions.} 
\label{fig2}
\end{figure*}

\begin{table}[]
\caption{Coding categories of SRL processes and definitions}
\label{table1}
\resizebox{1\textwidth}{!}{
\begin{tabular}{llll}
\hline
Main categories & SRL processes & Codes & Definitions \\ \hline
\multicolumn{1}{l|}{\multirow{4}{*}{Metacognition}} & Orientation & MCO & \begin{tabular}[c]{@{}l@{}}Orientation on the learning-related activities; \\ reading of general instruction and essay rubric.\end{tabular} \\ \cline{2-4} 
\multicolumn{1}{l|}{} & Planning & MCP & \begin{tabular}[c]{@{}l@{}}Planning of the reading and \\ writing process by arranging activities and determining strategies. \\ Proceeding to the next topic.\end{tabular} \\ \cline{2-4} 
\multicolumn{1}{l|}{} & Evaluation & MCE & \begin{tabular}[c]{@{}l@{}}Monitoring and checking the reading and writing process; \\ checking of progress according to instruction or plan.\end{tabular} \\ \cline{2-4} 
\multicolumn{1}{l|}{} & Monitoring & MCM & \begin{tabular}[c]{@{}l@{}}Evaluation of the learning process; \\ checking of content-wise correctness (e.g., the essay content) of learning activities.\end{tabular} \\ \hline
\multicolumn{1}{l|}{\multirow{2}{*}{Low Cognition}} & First-Reading & LCF & \begin{tabular}[c]{@{}l@{}}Reading information from the materials and pictorial \\ representations for the first time.\end{tabular} \\ \cline{2-4} 
\multicolumn{1}{l|}{} & Re-Reading & LCR & Rereading of information in the text or figures. \\ \hline
\multicolumn{1}{l|}{High Cognition} & Elaboration/Organization & HCEO & \begin{tabular}[c]{@{}l@{}}Elaborate by connecting content-related comments and concepts when writing the essay; \\ organizing content by creating an overview; \\ writing down information point by point in notes or essay window; summarizing; \\ adding information generated by oneself; \\ and editing information by rephrasing or integrating information with prior knowledge.\end{tabular} \\ \hline
\end{tabular}
}
\end{table}

\subsubsection{SRL process labeling}
Building on the approach proposed by \cite{fan2022towards} and adopted in several prior studies \citep{srivastava2022effects,li2024analytics,cheng2025self,zhao2025effect}, we began by translating trace events into identifiable learning actions using the action library, which labeled raw log data with meaningful behavioral indicators. The action library included a total of 17 learning actions, such as RELEVANT\_READING, WRITE\_ESSAY, and EDIT\_ANNOTATION. Next, following the theoretical framework by \cite{bannert2007metakognition}, we characterized the learning process in the multi-source writing task using three overarching categories: metacognition, low\_cognition, and high\_cognition. We defined seven corresponding SRL processes and mapped the learning actions to these processes via the process library, which identified SRL processes based on sequences of actions. Details of coding categories of SRL processes and definitions are presented in Table \ref{table1}. Given that the task involved writing with the assistance of ChatGPT (a GenAI tool), we added an additional label, CHATGPT, to the process library to capture interactions related to GenAI use.

\subsubsection{Data preprocessing}
Since our focus is on SRL strategies in GenAI-assisted writing, we included only students who used ChatGPT during their writing tasks, resulting in a sample of 165 students. Following previous studies \citep{srivastava2022effects,li2024analytics}, we then filtered out all 'NOT\_RECOGNIZED' and 'nan' entries in the process sequences to retain only informative data. To further refine the dataset, we removed outliers based on z-scores within three standard deviations \citep{colan2013and}, calculated from the length of the process sequences (i.e., process sequences that are too long or too short are removed.). This yielded a final dataset of 139 students, which we used for our study.

\subsubsection{RQ1 - Identifying hidden tactics}
We modeled the SRL process sequences using a HMM. A HMM is a probabilistic model in which observable events depend on a sequence of unobserved (hidden) states that follow a Markov process. This approach has been widely used in educational research for modeling sequential behaviors (e.g, writing behaviors) \citep{tadayon2020predicting, nguyen2024human,villalobos2024exploring,fincham2018study}.

As illustrated in Figure \ref{fig2} (the elements highlighted with a red background), $O$ denotes the observed sequences (i.e., SRL process sequences), while $S$ represents the hidden states (i.e., SRL hidden tactics). Because the hidden states $S$ cannot be directly observed, the objective is to infer them through the observed sequences $O$. Parameter estimation in a HMM is typically performed via maximum likelihood estimation. Two core components define a HMM: the transition matrix and the emission matrix. The transition matrix captures the probabilities of moving from one hidden state to another, whereas the emission matrix defines the probabilities of observing a particular event given a specific hidden state. To determine the optimal number of hidden states, we employed three widely-used criteria: the Akaike Information Criterion (AIC), the Bayesian Information Criterion (BIC), and Log-Likelihood (LL). BIC imposes a stronger penalty for model complexity and typically favors simpler models (fewer hidden states); lower values indicate better models. AIC also penalizes complexity but is more lenient, potentially selecting slightly more complex models; again, lower values are better. Log-Likelihood (LL) measures how well the model fits the data, with higher values indicating a better fit.

In our context, each observable SRL process corresponds to a SRL hidden tactic, suggesting that students were applying this tactic at that specific timestamp. These tactics manifest externally as observable SRL processes. Because hidden states represent unobservable internal tactics, we rely on the transition and emission matrices to uncover and interpret them.

\subsubsection{RQ2 - Identifying SRL strategies}
Building on previous research that identified SRL strategies based on SRL process sequences \citep{srivastava2022effects,li2024analytics,alghamdi2025analytics}, we applied clustering to identify SRL strategies as well, but based on sequences of hidden tactics rather than SRL processes. We used Levenshtein distance as the similarity measure for clustering sequences, as it effectively captures structural differences by accounting for insertions, deletions, and substitutions. To address the varying lengths of sequences, we applied a radial basis function to transform them into fixed-length similarity feature vectors. For the clustering process, we employed the K-means algorithm, a method widely adopted in previous educational research \citep{nguyen2024human, yang2024ink}. The optimal number of clusters was identified using the Elbow method \citep{bholowalia2014ebk}, based on two metrics: inertia, which reflects the sum of distances between data points and their corresponding centroids, and silhouette scores, which evaluate clustering performance by measuring how well each data point fits within its assigned cluster relative to others.

\subsubsection{RQ3 - Association with task performance}
To determine whether there were statistically significant differences in essay scores (i.e., task performance) among the identified SRL strategy groups, we first conducted a Shapiro-Wilk normality test, which indicated that the distribution of essay scores was not normal (W = 0.914, p = 2.298e-07). As a result, we used the non-parametric Mann-Whitney U test for the group comparison.

\subsubsection{Benchmark}
We also constructed a benchmark method for identifying SRL strategies without incorporating hidden tactics. The elements highlighted with a green background in Figure \ref{fig2} represent the processes specific to this benchmark method. All other components remain the same as in our proposed method, except for the inclusion of hidden tactics. This setup can also be viewed as an ablation study, demonstrating the validity of the proposed hidden tactics proxy in bridging SRL processes and SRL strategies. We first measured the similarity between the clustering results of both methods using three metrics: Homogeneity, Completeness, and V-Measure \citep{rosenberg2007v}. Homogeneity assesses whether each cluster in one result contains only members from a single cluster in the other result. Completeness checks whether all members of a given cluster in one result are assigned to the same cluster in the other result. V-Measure is the harmonic mean of homogeneity and completeness. All three metrics range from 0 to 1, with higher values indicating greater similarity. Next, we examined whether significant differences in task performance existed between the groups identified by both our method and the benchmark. We also examined the confusion matrix between the clustering results of the two methods and presented process distribution plots to provide further insights.

\section{RESULTS}
\subsection{Results on RQ1}

\begin{figure*}[hbt!]
\centering
\includegraphics[width=0.65\textwidth]{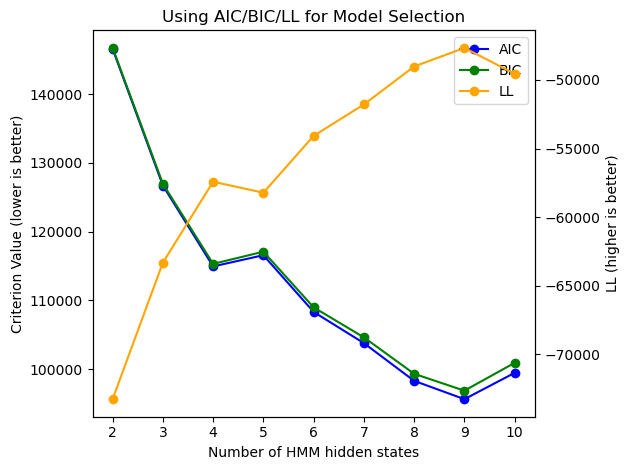}
\caption{Performance metrics by number of hidden states.} 
\label{fig3}
\end{figure*}

\begin{figure*}[hbt!]
\centering
\includegraphics[width=1\textwidth]{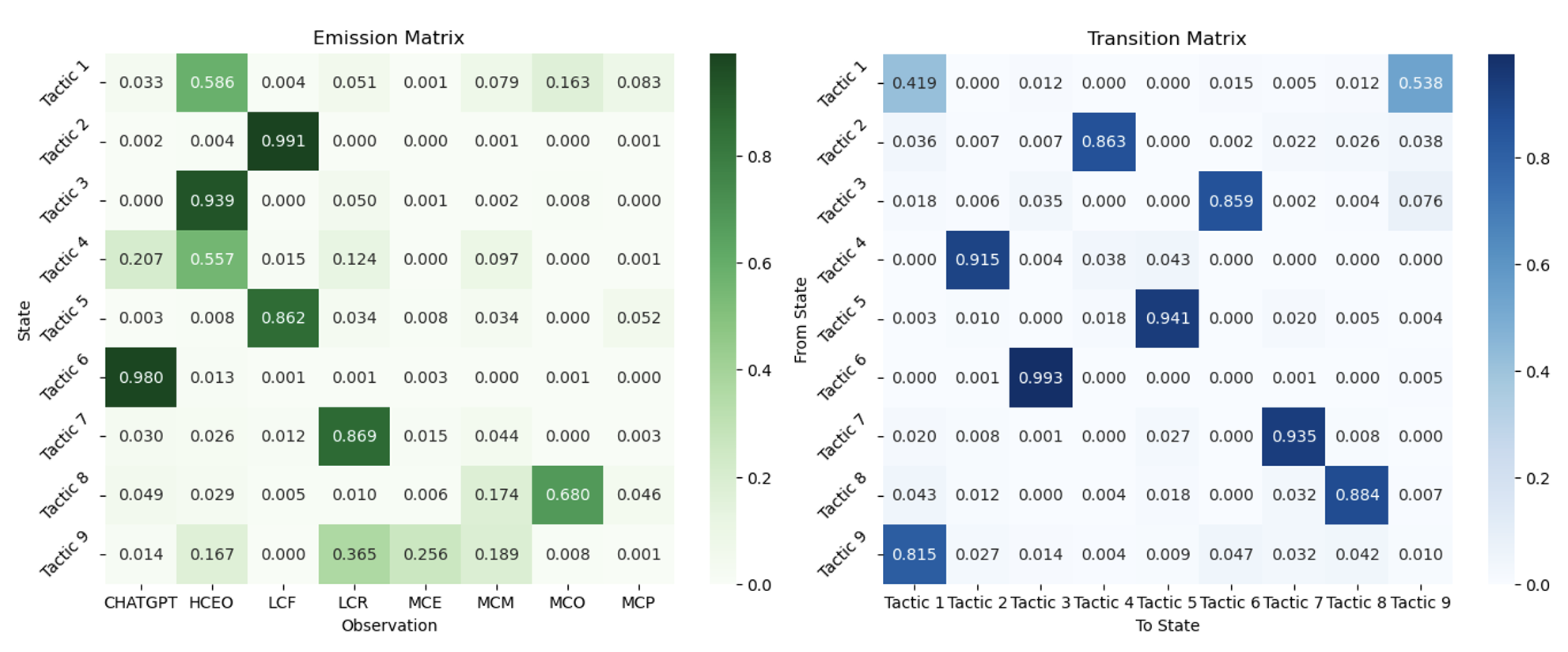}
\caption{Emission Matrix and Transition Matrix. } 
\label{fig4}
\end{figure*}

As shown in Figure \ref{fig3}, the AIC and BIC values were lowest, while the LL was highest when the number of hidden states was nine, indicating that the optimal number of SRL hidden tactics was nine. Figure \ref{fig4} provides a visual representation of the emission and transition matrices. Darker cells in emission matrix indicate that the corresponding observation is highly likely to occur when the system is in that specific hidden state. Darker cells in transition matrix indicate higher probabilities, meaning that a transition from the state on the y-axis to the state on the x-axis is more likely. Based on the emission and transition matrices, the hidden tactics can be interpreted as follows. 

\textbf{Hidden Tactic 1} had a high probability of emitting HECO (approximately 0.586) and MCO (approximately 0.163), suggesting that it reflects a tactic which students used to read the task requirements while writing. \textbf{Hidden Tactic 2} showed a high probability of emitting only LCF (about 0.991), indicating a tactic students primarily enacted in the initial reading of the materials. \textbf{Hidden Tactic 3} had a high probability of emitting only HCEO (approximately 0.939), suggesting a tactic students used when primarily focused on writing. \textbf{Hidden Tactic 4} exhibited a high probability of emitting a combination of CHATGPT (0.207), HCEO (0.557), and LCR (0.124), indicating a tactic students used to writing while engaging with GenAI, identifying gaps in their writing, and revisiting the materials to gather additional information. \textbf{Hidden tactic 5} had a high probability of emitting LCF (about 0.862), and lower probabilities for LCR (0.034), MCM (0.034), and MCP (0.052), indicating a tactic students primarily engage in initial reading, while also integrating some metacognitive monitoring and planning activities. \textbf{Hidden Tactic 6} showed a strong likelihood of emitting only CHATGPT (approximately 0.98), indicating a tactic students primarily used to interact with the GenAI tool. \textbf{Hidden Tactic 7} was highly associated with LCR alone (approximately 0.869), suggesting a focus on re-reading learning materials. \textbf{Hidden Tactic 8} displayed a notable combination of MCM (0.174) and MCO (0.68), reflecting a tactic centered on understanding task requirements while evaluating the learning process. \textbf{Hidden Tactic 9} involved a mix of HECO (0.167), LCR (0.365), MCE (0.256), and MCM (0.189), pointing to a complex tactic students engaged to write, evaluate and monitor their learning processes and products, as well as to revisit materials to verify information. 

Based on the identified hidden tactics, we found that while some tactics were solely related to a single primary SRL process (i.e., hidden tactics 2, 3, 6, and 7), others (i.e., hidden tactics 1, 4, 5, 8, and 9) captured multiple intertwined SRL processes. Since the hidden tactics were interpreted through the emission probabilities, they also accounted for some noise and outlier situations. For example, although hidden tactic 6 primarily focused on interaction with the GenAI tool, it still assigned a total probability of 0.02 to all other SRL processes.

Based on the transition matrix, we can also observe the transitions between the hidden tactics students were using. We identified some persistent hidden tactics (i.e., remaining within the same tactic) as well as some high-probability transitions (i.e., shifting between tactics). After engaging with \textbf{Hidden Tactic 1} (writing while reading task requirements), students often continued using the same tactic. However, there was also a strong tendency to shift to \textbf{Hidden Tactic 9}, where they engage in more complex self-regulation and writing activities. From \textbf{Hidden Tactic 2} (initial reading), students frequently transitioned to \textbf{Hidden tactic 4}, indicating a move into writing while using GenAI and revisiting the materials. When students were using \textbf{Hidden Tactic 3} (focused writing), they tended to shift into  \textbf{Hidden Tactic 6}, where they primarily interacted with GenAI; this suggests they might have sought support or feedback from GenAI. After \textbf{Hidden Tactic 4} (writing with GenAI and identifying gaps), students often circled back to \textbf{Hidden Tactic 2}, resuming initial reading to find additional information. \textbf{Hidden Tactic 5} (initial reading with some metacognitive activities) was relatively stable, with students typically remaining in that tactic. Following \textbf{Hidden Tactic 6} (heavy GenAI use), students usually transitioned to \textbf{Hidden Tactic 3}, returning to focused writing after engaging with the GenAI tool. Students in \textbf{Hidden Tactic 7} (re-reading materials) also tended to persist in that tactic. Similarly, \textbf{Hidden Tactic 8} (understanding the task while evaluating their process) showed high persistence. Finally, after \textbf{Hidden Tactic 9} (complex evaluation, monitoring, and writing), students frequently returned to \textbf{Hidden Tactic 1}, resuming their task with a focus on writing while referencing requirements.

\subsection{Results on RQ2}
\begin{figure*}[hbt!]
\centering
\includegraphics[width=0.65\textwidth]{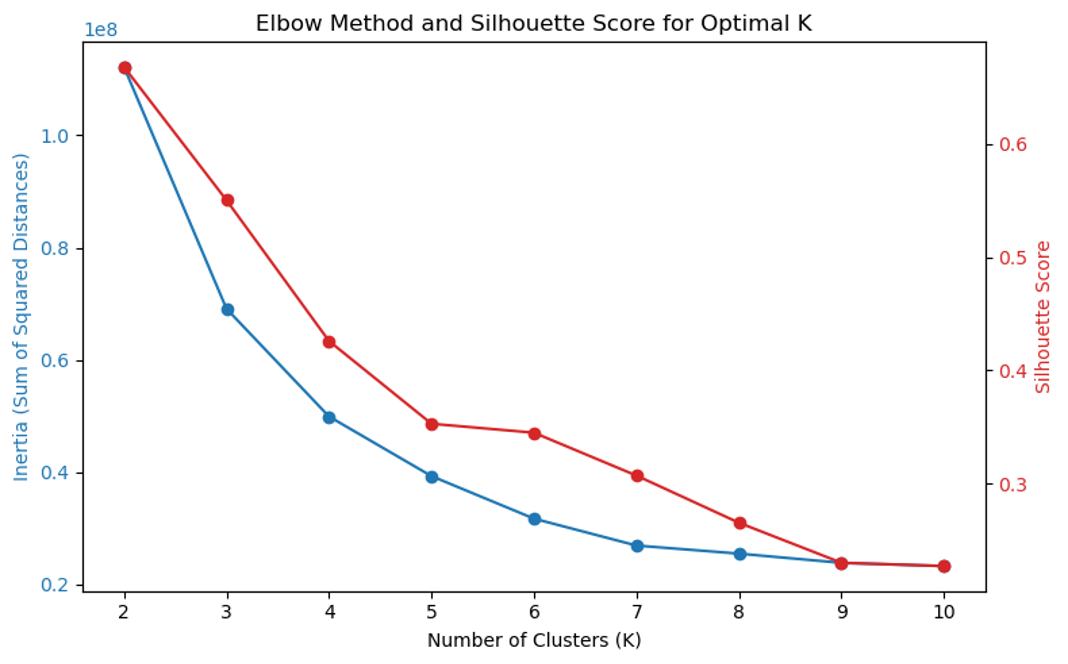}
\caption{Elbow plot.} 
\label{fig5}
\end{figure*}

\begin{figure*}[hbt!]
\centering
\includegraphics[width=1\textwidth]{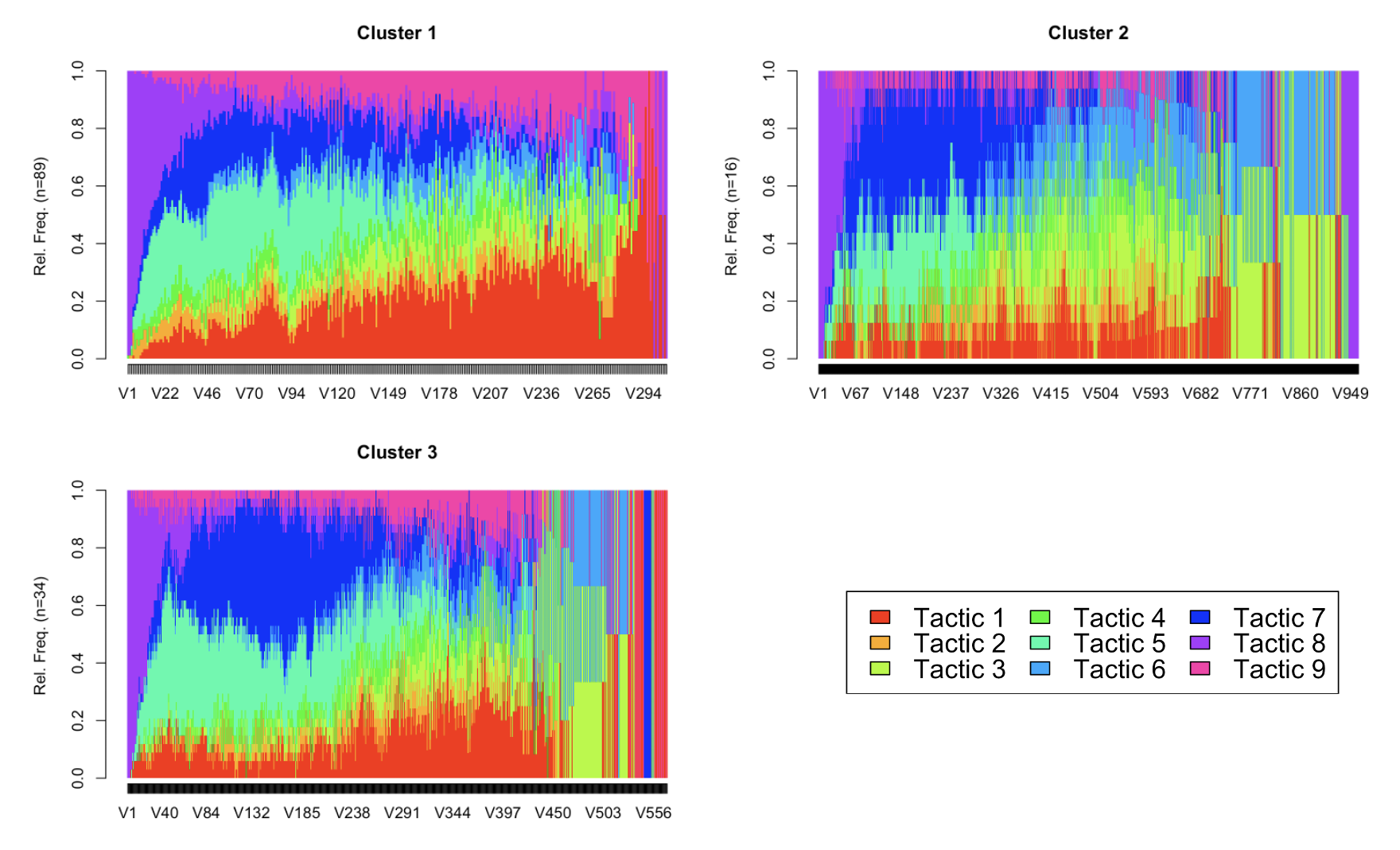}
\caption{Hidden tactics distribution plot.} 
\label{fig6}
\end{figure*}

\begin{figure*}[hbt!]
\centering
\includegraphics[width=0.65\textwidth]{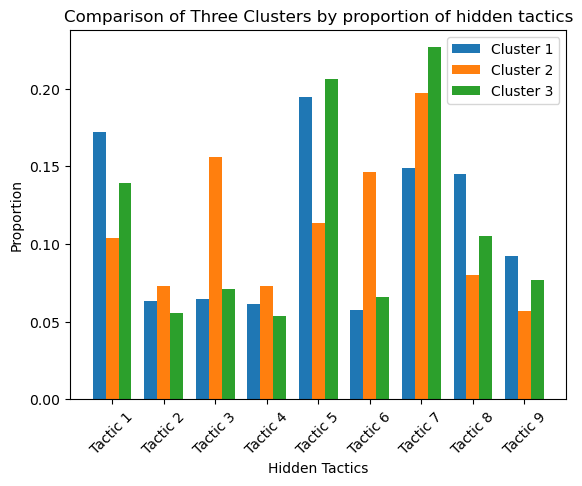}
\caption{Proportions of hidden tactics.} 
\label{fig7}
\end{figure*}

The Elbow plot shown in Figure \ref{fig5} indicates that the optimal number of clusters is three, with a silhouette score of approximately 0.55, suggesting a reasonably good clustering quality. We successfully identified three clusters, each representing a distinct SRL strategy based on the hidden tactics students employed during the writing tasks. The distribution plots of hidden tactics for each cluster are presented in Figure \ref{fig6}, while the proportion of hidden tactics within each cluster is shown in Figure \ref{fig7}. We interpreted the SRL strategies based on insights from these two figures.

\noindent \textbf{Cluster 1: Conventional Strategic Writers.} Students in this cluster would begin their tasks by heavily engaging with Hidden Tactic 8, focusing primarily on understanding task requirements and evaluating their learning processes. They then transitioned to Hidden Tactic 5, engaging mostly in the initial reading of materials along with some metacognitive planning. They also occasionally made use of Hidden Tactic 7 (re-reading materials). As they moved into the later phases of their tasks, their attention notably shifted towards Hidden Tactic 1 (writing while simultaneously consulting task requirements) and Hidden Tactic 9 (complex interaction involving writing, self-evaluation, monitoring, and revisiting materials). Overall, this group demonstrated a conventional and structured approach: carefully reading task requirements first, reviewing the provided materials, and finally writing their responses by systematically integrating insights from both task requirements and study materials. Given the proportion of each hidden tactic within the cluster, their strong engagement in Hidden Tactics 1, 8, and 9 suggests a methodical progression from reading to synthesis. Conversely, their low engagement in Hidden Tactics 3, 6, and 7 indicates minimal reliance on unplanned writing, the GenAI tool, or repetitive material review.

\noindent \textbf{Cluster 2: GenAI-Integrated Writers.} Students in this cluster maintained consistent engagement with Hidden Tactic 1 throughout all phases, consistently writing while actively consulting task requirements. Initially, they briefly engaged with Hidden Tactic 8 (understanding task requirements and evaluating the learning process) but rapidly transitioned to Hidden Tactic 7 (re-reading materials) and Hidden Tactic 5 (initial reading with metacognitive monitoring). As their work on the task progressed, students strongly shifted their attention toward Hidden Tactic 3 (writing), Hidden Tactic 4 (writing with GenAI assistance, identifying gaps, and revisiting materials), and Hidden Tactic 6 (exclusive interaction with GenAI tools). This cluster distinctly integrated GenAI into their writing process, preferring an active writing approach combined with frequent interactions with the GenAI tool. Given the proportion of each hidden tactic, this cluster heavily utilized GenAI and engaged in continuous writing and revision, as evidenced by high proportion in Hidden Tactics 2, 3, 4, and 6. Their lower engagement in Hidden Tactics 1, 5, 8, and 9 highlighted a tendency to de-prioritize initial task review, deep reading, and manual synthesis in favor of requesting AI supports.

\noindent \textbf{Cluster 3: Intensive Material Reviewers.} Students in this cluster initially focused predominantly on Hidden Tactic 8, much like Cluster 1, emphasizing task requirements and learning evaluation. In the subsequent phases, their attention was strongly directed toward Tactic 5, with significant engagement in initial material reading combined with metacognitive monitoring and planning, accompanied by a noteworthy investment in Hidden Tactic 7 (re-reading). Notably, this middle phase extended longer for this group compared to the other two clusters, reflecting potential difficulties in fully grasping the study materials. In their later phases, students shifted attention toward Hidden Tactic 1 (writing alongside task requirements) and Hidden Tactic 9 (complex engagement involving writing, evaluation, and revisiting materials). This cluster was characterized by intensive efforts in comprehending materials, resulting in considerable re-reading, suggesting that students in this group likely encountered challenges in content comprehension or synthesis. Given the proportion of each hidden tactic, this cluster was defined by intensive engagement with material comprehension and verification, demonstrated by their strong preference for Hidden Tactics 5 and 7. Their low engagement in Hidden Tactics 4 underscored a limited use of GenAI for support or guidance.

\subsection{Results on RQ3}
\begin{table}[]
\caption{Statistics of essay scores within the clusters. Bolded values indicate statistically significant results (p < 0.05). }
\label{table2}
\centering
\begin{tabular}{@{}cccc@{}}
\toprule
 & \begin{tabular}[c]{@{}c@{}}Essay Scores\\ Mean (Standard Deviation)\end{tabular} & \begin{tabular}[c]{@{}c@{}}Mann-Whitney U Test\\ Statistic (P-value)\end{tabular} & \begin{tabular}[c]{@{}c@{}}Effect Size\\ Absolute Cohen's r\end{tabular} \\ \midrule
Cluster 1 (n = 89) & 16.674 (3.637) & \textbf{\begin{tabular}[c]{@{}c@{}}vs. Cluster 2 \\ 481.5 (0.039)\end{tabular}} & 0.324 \\ \midrule
Cluster 2 (n = 16) & 18.438 (2.549) & \textbf{\begin{tabular}[c]{@{}c@{}}vs. Cluster 3 \\ 402.0 (0.006)\end{tabular}} & 0.478 \\ \midrule
Cluster 3 (n = 34) & 15.735 (4.082) & \begin{tabular}[c]{@{}c@{}}vs. Cluster 1 \\ 1726.5 (0.226)\end{tabular} & 0.141 \\ \bottomrule
\end{tabular}
\end{table}

The results of the statistical tests examining differences in essay scores among clusters (i.e., students grouped by their SRL strategy use)  were presented in Table \ref{table2}. We found that task performance in Cluster 2 (GenAI-Integrated Writers) was significantly higher than that of both Cluster 1 (Traditional Strategic Writers) (statistic = 481.5, p < 0.05, r = 0.324) and Cluster 3 (Intensive Material Reviewers) (statistic = 402.0, p < 0.05, r = 0.478). These findings align with previous research suggesting that GenAI can enhance students' task performance in final written work \citep{yang2025modifying}. The lowest mean scores observed in Cluster 3 further support our interpretation that writers in this group invested substantial effort in understanding the materials, involving extensive re-reading. This suggests that students in this group likely encountered challenges in content comprehension or synthesis.
\subsubsection{Benchmark Comparison}

\begin{figure*}[hbt!]
\centering
\includegraphics[width=0.65\textwidth]{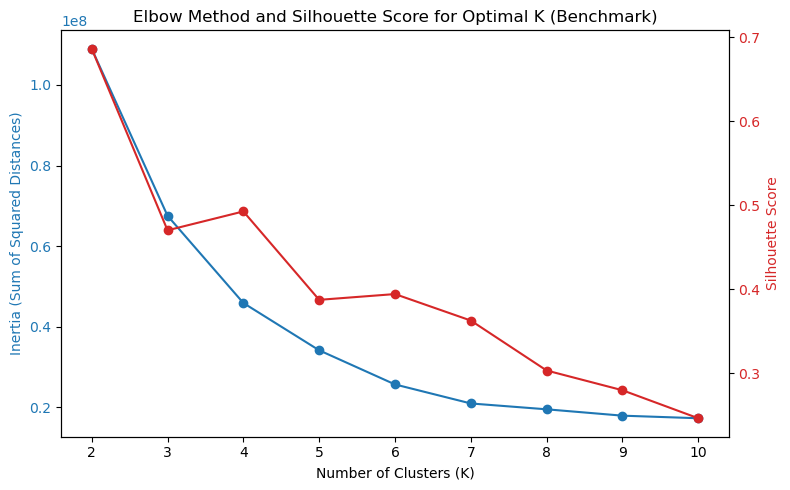}
\caption{Elbow plot for benchmark method.} 
\label{fig8}
\end{figure*}

\begin{table}[]
\caption{Statistics of essay scores within the clusters (Benchmark)}
\label{table3}
\centering
\resizebox{1\textwidth}{!}{
\begin{tabular}{@{}cccccc@{}}
\toprule
{} & \begin{tabular}[c]{@{}c@{}}Essay Scores\\ Mean (Standard Deviation)\end{tabular} & \begin{tabular}[c]{@{}c@{}}Mann-Whitney U Test\\ Statistic (P-value) vs. Cluster 1\\ (Benchmark)\end{tabular} & \begin{tabular}[c]{@{}c@{}}Mann-Whitney U Test\\ Statistic (P-value) vs. Cluster 2\\ (Benchmark)\end{tabular} & \begin{tabular}[c]{@{}c@{}}Mann-Whitney U Test\\ Statistic (P-value) vs. Cluster 3\\ (Benchmark)\end{tabular} & \begin{tabular}[c]{@{}c@{}}Mann-Whitney U Test\\ Statistic (P-value) vs. Cluster 4\\ (Benchmark)\end{tabular} \\ \midrule
\begin{tabular}[c]{@{}c@{}}Cluster 1\\ (Benchmark)\\ (n=19)\end{tabular} & 16.526 (2.663) & - & 652.0 (0.691) & 50.5 (0.087) & 340.5 (0.733) \\
\begin{tabular}[c]{@{}c@{}}Cluster 2\\ (Benchmark)\\ (n=73)\end{tabular} & 16.658 (3.786) & - & - & 230.0 (0.144) & 1403.5 (0.920) \\
\begin{tabular}[c]{@{}c@{}}Cluster 3\\ (Benchmark)\\ (n=9)\end{tabular} & 18.333 (2.828) & - & - & - & 232.5 (0.096) \\
\begin{tabular}[c]{@{}c@{}}Cluster 4\\ (Benchmark)\\ (n=38)\end{tabular} & 16.289 (4.116) & - & - & - & - \\ \bottomrule
\end{tabular}
}
\end{table}

The Elbow plot for benchmark method (i.e. without hidden tactics) shown in Figure \ref{fig8} indicates that the optimal number of clusters was four, with a silhouette score of approximately 0.5, suggesting a reasonably good clustering quality. The results of the statistical tests among clusters identified by the benchmark method are presented in Table \ref{table3}. However, no significant differences in task performance were found among the clusters. This further demonstrates that the use of hidden tactics is capable of accurately characterizing the different SRL strategies adopted by students and the subsequent impact on their writing performance. We also compared the clustering results of our proposed method with those of the benchmark. The results -- Homogeneity: 0.685, Completeness: 0.527, V-Measure: 0.596 -- indicate a moderate similarity between the two approaches, further supporting that our proposed method can identify SRL strategies that previous benchmark methods could not. 

\begin{figure*}[hbt!]
\centering
\includegraphics[width=1\textwidth]{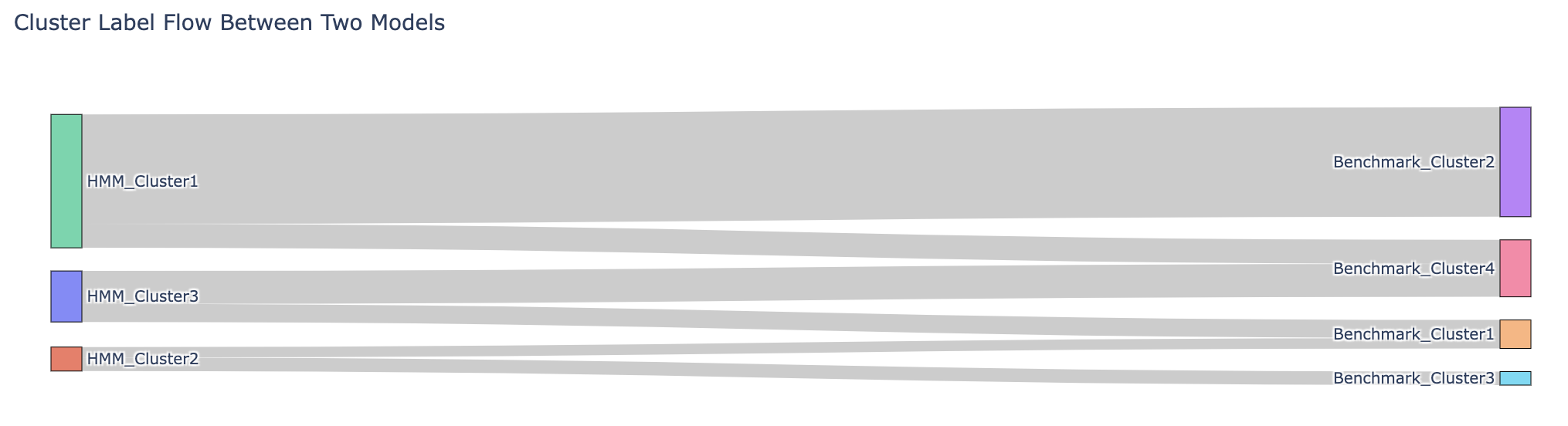}
\caption{Sankey diagram between clustering results from proposed method and benchmark method.}
\label{fig9}
\end{figure*}

\begin{figure*}[hbt!]
\centering
\includegraphics[width=1\textwidth]{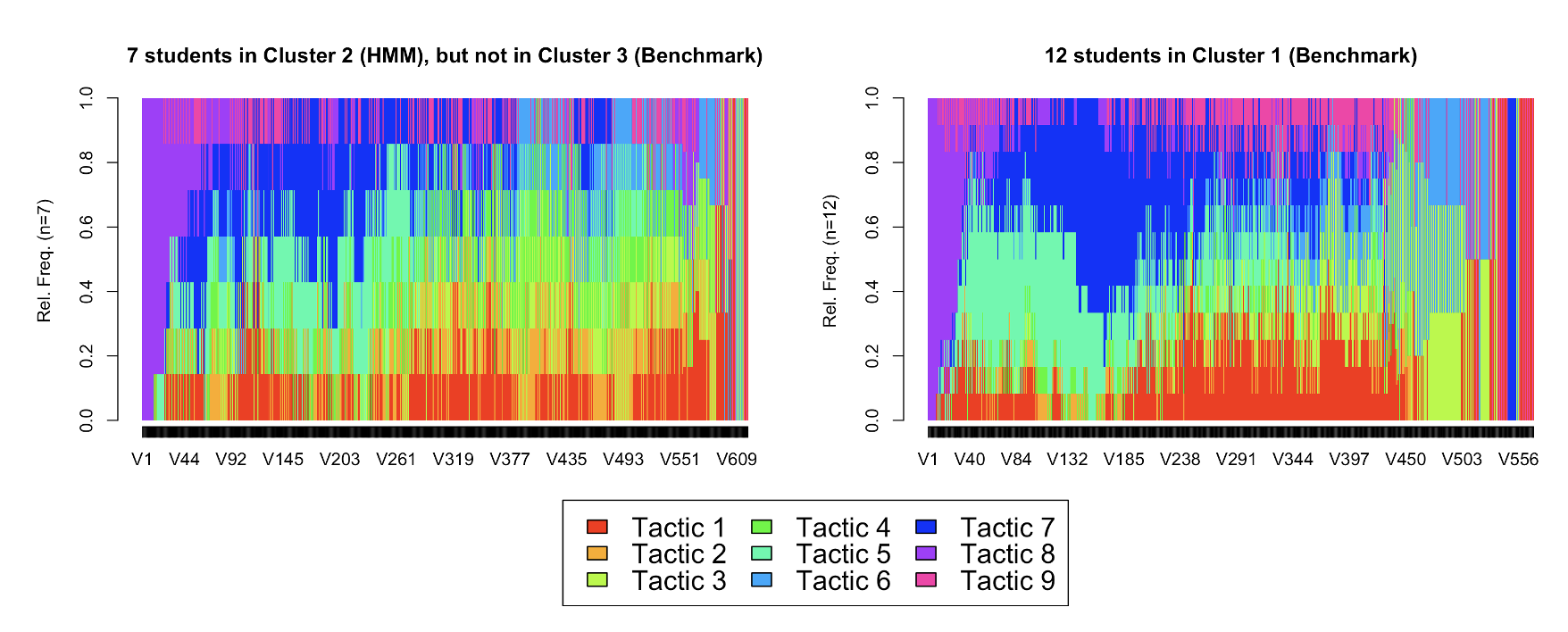}
\caption{7 students vs 12 students by hidden tactics distribution.}
\label{fig10}
\end{figure*}

\begin{figure*}[hbt!]
\centering
\includegraphics[width=1\textwidth]{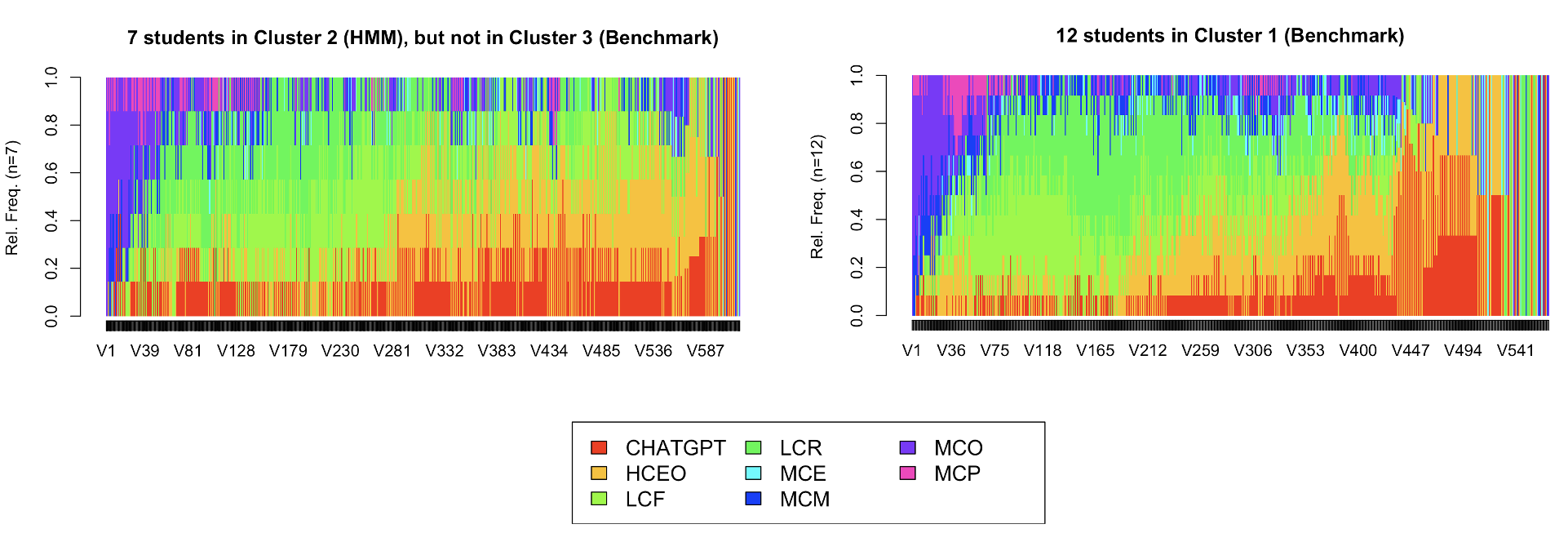}
\caption{7 students vs 12 students by SRL processes distribution}
\label{fig11}
\end{figure*}

To further compare the clustering results of the two methods, we present the contingency table in Table \ref{table4} and the Sankey diagram in Figure \ref{fig9}. Interestingly, the high-performance group (i.e., Cluster 2 identified with the use of hidden tactics) included 7 students who were not identified by the benchmark method’s high-performance group (i.e., Cluster 3 in the Benchmark method). These 7 students had a mean score of 18.571 (SD = 2.299). In contrast, the benchmark method placed them in Cluster 1 (Benchmark) alongside 12 other students, whose mean score was lower at 15.333 (SD = 2.269). As a result, we conducted a further comparison of the hidden tactic sequences and SRL processes sequences between the groups of these 7 and 12 students in Cluster 1 detected by the Benchmark method to understand why this discrepancy occurred. The results are shown in Figures \ref{fig10} and \ref{fig11} and indicate that the two groups of students had very different distributions of hidden tactics but very similar distributions of SRL processes. This explains why methods based on hidden tactics can differentiate between the groups, while methods based on SRL processes alone cannot. This case analysis further demonstrated that the identified SRL strategies through hidden tactics captured more information than those identified through SRL processes.

Overall, identifying SRL strategies through hidden tactics produces different results compared to identification through SRL processes alone. Notably, task performance was only significantly different among SRL strategies identified by the proposed method that was based on the hidden tactic detection. Moreover, hidden tactics revealed distinctions among groups of students that SRL processes alone failed to capture. These findings collectively demonstrate the validity of the proposed method based on hidden tactic detection compared to the benchmark method that used SRL processes alone.

\begin{table}[]
\caption{Contingency table between clustering results from proposed method and benchmark method.}
\label{table4}
\centering
\begin{tabular}{@{}ccccc@{}}
\toprule
{} & \begin{tabular}[c]{@{}c@{}}Cluster 1\\ (Benchmark)\end{tabular} & \begin{tabular}[c]{@{}c@{}}Cluster 2\\ (Benchmark)\end{tabular} & \begin{tabular}[c]{@{}c@{}}Cluster 3\\ (Benchmark)\end{tabular} & \begin{tabular}[c]{@{}c@{}}Cluster 4\\ (Benchmark)\end{tabular} \\ \midrule
\begin{tabular}[c]{@{}c@{}}Cluster 1\\ (HMM)\end{tabular} & 0 & 73 & 0 & 16 \\
\begin{tabular}[c]{@{}c@{}}Cluster 2\\ (HMM)\end{tabular} & 7 & 0 & 9 & 0 \\
\begin{tabular}[c]{@{}c@{}}Cluster 3\\ (HMM)\end{tabular} & 12 & 0 & 0 & 22 \\ \bottomrule
\end{tabular}
\end{table}

\section{DISCUSSION AND CONCLUSIONS}
In this study, we aimed to bridge the gap in current modeling approaches to SRL strategies, as trace-based analytical methods often oversimplify SRL processes and fail to capture their discontinuous, non-linear, and intertwined nature. We proposed that latent states (i.e., hidden tactics) exist between SRL processes and SRL strategies, providing a more valid representation of SRL. To investigate the SRL strategies learners employ during writing tasks involving GenAI, we used a HMM to identify these hidden tactics in a dataset of higher education students who utilized a GenAI tool as part of their writing tasks. We then clustered the sequences of hidden tactics to identify SRL strategies and examined their associations with task performance. To further demonstrate the validity of our method, we benchmarked it against traditional approaches that model SRL strategies directly from SRL processes, comparing the results to those of our proposed method.

In response to RQ1, we identified nine distinct hidden tactics within the dataset. Notably, several of these hidden tactics encompass multiple SRL processes, underscoring the complex nature of students' learning behaviors. For instance, Hidden Tactic 9 is characterized by a blend of SRL processes, including HECO, LCR, MCE, and MCM. This combination reveals a sophisticated pattern of engagement through which students were involved in writing, self-regulated evaluation, monitoring, as well as revisiting materials to verify information. This finding reinforces the idea that SRL processes do not occur in isolation or follow a linear progression. Rather, they are discontinuous, non-linear, and deeply intertwined, aligning with insights from previous research on the complexity of SRL processes \citep{dever2023complex,li2020examining}. 

In response to RQ2, we identified three distinct types of SRL strategies: \textbf{Conventional Strategic Writers}, \textbf{GenAI-Integrated Writers}, and \textbf{Intensive Material Reviewers}. Compared to SRL strategies reported in previous studies \cite{srivastava2022effects,li2024analytics}, which primarily describe the general sequence of SRL processes (e.g., “Read First, Write Next”), our findings offer a more nuanced perspective by highlighting the presence of hidden tactics. For instance, although multiple hidden tactics involve writing, they differ in subtle but important ways. \textbf{Hidden Tactic 3} was primarily focused on writing alone, whereas \textbf{Hidden Tactic 4} involved writing in conjunction with GenAI tools. \textbf{Hidden Tactic 1}, on the other hand, represents reading task requirements with writing without involving much GenAI. These varied approaches to writing reflect the diverse ways learners regulate their learning processes. For example, \textbf{GenAI-Integrated Writers} predominantly utilized \textbf{Hidden Tactic 4}, while \textbf{Conventional Strategic Writers} were more aligned with \textbf{Hidden Tactic 1}. As a result, the proposed modeling approach provides a more fine-grained understanding of SRL strategies, enabling the capture of the intertwined relationships between SRL processes when identifying SRL strategies — relationships that were not captured in previous studies \citep{srivastava2022effects,li2024analytics}.

In response to RQ3, we found that \textbf{GenAI-Integrated Writers} demonstrated significantly higher task performance compared to the other two groups. This finding aligns with previous research showing that integrating GenAI tools can enhance writing performance, particularly in terms of essay quality \citep{yang2025modifying}. In contrast, \textbf{Intensive Material Reviewers} had the lowest average task performance. This may be attributed to the considerable amount of time they spent reading, which aligns with earlier studies on time allocation in timed writing tasks \citep{tabari2017investigating}, suggesting that students who devote excessive time to reading or planning may struggle to complete or revise their essays effectively, ultimately leading to lower scores. Additionally, we found no significant differences in task performance across the benchmark method's clustering results. This is likely because clustering based solely on SRL processes failed to distinguish certain high-performing students that our method successfully captures. This finding echoes our results from RQ2, suggesting that even when SRL processes appeared similar, the underlying tactics could differ, leading to variations in performance.

\subsection{Implications}
According to our results, \textbf{GenAI-Integrated Writers} demonstrated significantly higher task performance compared to the other two groups. However, we also found that \textbf{GenAI-Integrated Writers} engaged less in Hidden Tactic 9 (a complex interaction involving writing, self-evaluation, monitoring, and revisiting materials) and more in Hidden Tactic 6 (exclusive interaction with GenAI tools). This finding aligns with the recent debate on performance versus learning confusion with GenAI \citep{weidlich2025chatgpt}, where it was argued that some measures of performance are not sufficient to demonstrate true learning. Our findings are also consistent with previous studies highlighting the metacognitive laziness (i.e., hinder stduents' ability to self-regulate and engage deeply in learning.) introduced by GenAI \citep{fan2025beware} and the over-reliance on GenAI tools \citep{chen2025unpacking}. Although students showed high task performance, they exhibited low engagement in meaningful cognitive and metacognitive processes, instead focusing more on interactions with GenAI. This highlights the need for researchers and practitioners to be aware of students' potential over-reliance on GenAI techniques, which may lead them to miss valuable opportunities for deep, meaningful learning.

\subsection{Limitations and future work}
We acknowledge that our study has several limitations. First, our analysis was conducted using only one writing dataset. This may limit the generalization of our results, as the characteristics of this specific dataset may not represent a broader range of writing tasks or educational contexts. Future studies should replicate this approach across diverse datasets to validate its broader applicability. Second, we broadly labeled all interactions with GenAI as "CHATGPT," without accounting for the diverse ways students might engage with the tool. In reality, students may vary significantly in how they prompt, interpret, and incorporate AI-generated suggestions, which could influence the outcomes. Future research should investigate these interaction patterns in greater depth (potentially through the analysis of chat interactions and content), in order to develop a more diverse and nuanced set of code labels that reflect the varied ways students engage with GenAI. Third, we did not collect think-aloud data to support the criterion-related validity of the proposed method. The absence of such data makes it difficult to verify the SRL strategies underlying students’ interactions with the digital platform, potentially limiting our understanding of how the method aligns with actual user learning strategies.

\bibliography{sample}
\end{document}